# Remote dispersion scan: transformer-network retrieval of ultrafast pulses after non-linear propagation

KEVIN WATSON[1], YUTONG GENG[1], TOBIAS SAULE[1], THOMAS WEINACHT[2], AND CARLOS A. TRALLERO-HERRERO[1,*]

[1] *Department of Physics, University of Connecticut, Storrs, CT 06269, USA*
[2] *Department of Physics and Astronomy, Stony Brook University, Stony Brook, New York 11794, USA*
[*] *Corresponding author: carlos.trallero@uconn.edu*

**Abstract:** Accurate and rapid characterization of broadband electric fields is essential for all ultrafast applications and remains an active field of research. In this work, we introduce remote dispersion-scan, a transformer-neural-network enabled dispersion-scan–based pulse-characterization method that can characterize femtosecond laser pulses. A "local" scan of the non-linear spectral phase before several linear and non-linear processes, including amplification, compression, and self-phase modulation, allows for the field retrieval "remotely" at the interaction region. We show that the reconstruction accuracy obtained from a single measurement of the fundamental and second harmonic is comparable to that of a full two-dimensional scan. We confirm the technique experimentally by compressing a 300 W, 1.3 ps, 1030 nm pulse in a hollow core fiber to 100 fs and measuring the fundamental and second harmonic spectra while scanning the second order phase in a pulse shaper before power amplification. These results establish a simple, robust, alignment insensitive live-view pulse reconstruction modality.

## 1. Introduction

High-power ultrafast laser sources have become indispensable tools in both fundamental science and industry [1–3]. To maximize both peak and average power, many laser systems are transitioning from broadband Ti:Sapphire gain media to high–average power Yb-doped architectures [4–7]. The superior average-power scaling of Yb-based systems, however, comes at the expense of usable bandwidth [8]; achieving the shortest pulse durations and highest peak intensities therefore requires secondary spectral-broadening stages [5, 9], for which self-phase modulation (SPM) in a hollow core fiber (HCF) or multipass cell is the conventional method [10–13]. SPM dynamically reshapes a pulse's spectrum, coupling it to phase through a complex, non-linear relationship [14]. Neural networks—particularly transformer architectures that have driven recent advances in artificial intelligence [15–17]—have demonstrated exceptional capability in learning such complex mappings [18–24]. In fact, deep learning has demonstrated the ability to compensate for non-linear effects in fibers [25–27] and to retrieve arbitrary ultrafast pulses from pairs of pre- and post-SPM spectra [28]. This is very relevant to high average and peak power laser systems, where non-linear propagation is non-trivial.

In this work, we introduce the remote-dispersion-scan (RD-scan) approach, in which dispersion is varied prior to the chirped pulse amplification (CPA) step for pulses that undergo >10x compression from 1.3 ps to 100 fs at 300 W for a 1030 nm Yb laser. A transformer neural network (NN) is trained to recover pulses by learning the non-linear mapping between remotely scanned SPM spectra and the underlying spectral phase and pulse shape. We empirically demonstrate that a small set of measured spectra—capturing both the fundamental and second-harmonic components after SPM broadening and compression—contains sufficient information for accurate pulse reconstruction on both simulated and experimental measurements. Fast convergence arises from the high quality of the training set. Indeed, the RD-scan NN is trained on an expansive ensemble of ultrafast pulses simulated with the state-of-the-art non-linear propagation package `Luna.jl`, enabling robust generalization over a nine-dimensional parameter space. By systematically evaluating RD-scan retrieval accuracy as a function of dispersion-scan length, we show that accurate pulse characterization is achievable with a single measurement of the fundamental and second harmonic spectra, and experimentally demonstrate live-view pulse retrieval.

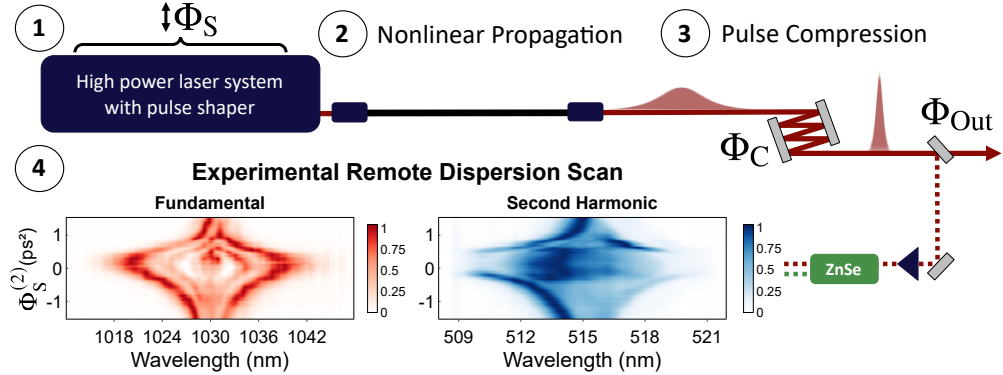

Fig. 1. **Experimental layout of remote dispersion scan:** The RD-scan methodology is given in four steps. **(1)** Laser system with integrated pulse shaper. **(2)** Hollow core fiber broadening. **(3)** Pulse compression by a series of chirped mirrors. **(4)** Spectral measurement of the pulse with SHG.

## 2. Methods

The RD-scan experimental layout is shown in Fig. 1. The full laser system is detailed in [29] and shown in greater detail in supplementary section 1. Briefly, we use an Amphos Yb laser producing 1030 nm, 1.3 ps pulses at 300 W average power with a tunable repetition rate from 25 kHz to 1 MHz while maintaining constant power. Pulses are compressed >10-fold by broadening in a stretched HCF filled with Kr, followed by chirped mirrors. A low-power pulse shaper [30, 31], commonly integrated in high power lasers prior to the main amplification stage [32–37], is used to apply a quadratic spectral phase, $\Phi_S^{(2)}$ or GDD at (1). Following amplification, pulses undergo non-linear spectral broadening via self-phase modulation (SPM) in a stretched HCF (2) [10, 12, 38–40], and are subsequently compressed using chirped mirrors (3) [41]. At the point of measurement, the high–peak-power beam is sampled, and second-harmonic light is generated using a random quasi-phase-matching non-linear crystal (ZnSe in this work) [42, 43] for spectrometer-based measurement (4). The $\Phi_S^{(2)}$ scanned fundamental and second-harmonic spectral components are then used to recover the phase of the electric field for the sampled output pulses, $\Phi_{\text{Out}}$.

Although similar in methodology to a conventional dispersion scan [44–47], our approach offers a key advantage: both the fundamental and second-harmonic spectra evolve continuously during the dispersion scan as a consequence of dynamically altering SPM in the fiber. These spectrally rich traces encode direct signatures of non-linear pulse propagation which substantially increase the information content available to the model. As a result, the transformer network can accurately infer pulse characteristics (for example spectral phase) at multiple points along the propagation axis, enabling robust reconstruction and diagnostics. Further, the scanning is performed remotely and programmatically; thus, there is no need for a dedicated scanning setup at the point of measurement, as non-linear propagation is more of an effective mechanism for pulse characterization rather than an unwanted complication.

The instantaneous frequency shift due to SPM is given by:

$$\Delta\omega(t) = -\frac{d\Phi_{\text{NL}}(t)}{dt} = -k_0\, n_2\, L\, \frac{dI(t)}{dt} \quad (1)$$

This couples the intensity profile of a pulse, $I(t)$, to the resultant spectral broadening $\Delta\omega(t)$ of SPM over non-linear propagation [48]. Thus, SPM encodes spectral phase information into the fundamental spectrum. However, the fundamental is not coupled to phase effects that occur after SPM (which are predominantly second order such as pulse compression and GDD). Therefore we rely on the second harmonic to detect such phase effects. This additional information is particularly important for high average and peak power systems, where non-linear propagation effects are prevalent. Together, the fundamental and second harmonic are embedded with phase information from multiple points along propagation, including from the laser system ($\Phi_S$), non-linear propagation ($\Phi_{NL}$), compression ($\Phi_C$), and the point of measurement ($\Phi_{Out}$), yielding rich spectral features for a predictive network to decode for pulse recovery.

This RD-scan transformer NN was trained on simulated data, enabling exhaustive and consistent sampling to a degree that is not experimentally feasible. [49]. Fiber propagation was simulated with `Luna.jl` [50], a state-of-the-art non-linear pulse-propagation package tailored for hollow-core capillaries and fibers. `Luna.jl` implements the unidirectional pulse propagation equation (and related Generalized Non-Linear Schroedinger Equation models) [48, 51, 52], accounting for loss, ionization, multimode propagation, Kerr non-linearities, and their consequent effects (e.g., SPM and self-steepening).

Over 5.1 million unique simulated pulses were used to train the RD-scan transformer across a nine-dimensional parameter space encompassing variations in pulse (pulse energy $\mathcal{E}$, phase orders [53] $\Phi_S^{(2)}$, $\Phi_S^{(3)}$, $\Phi_S^{(4)}$), fiber (gas pressure $p$, length $l$, core diameter $d$), and phase compression ($\Phi_C^{(2)}$, $\Phi_C^{(3)}$). In order to ensure robust retrieval performance, we explicitly incorporate variations in fiber geometry and compression into the simulation ensemble. Our laser operates over a broad repetition-rate range (25 kHz–1 MHz), which modifies the optimal fiber pressure and compression; consequently, these degrees of freedom are sampled when generating the training set for the RD-scan transformer. By parameterizing hollow-core fiber length and diameter alongside compression and pressure, a single network achieves consistent performance across instrument operating modes and across fibers that are exchanged on a semi-regular basis. Expanding the training manifold thus promotes learning of the underlying physical mappings relevant to the retrieval task, rather than spurious, regime-specific correlations [54]. All simulations were constrained to experimentally accessible operating conditions in order to preserve physical consistency. In practice, the parameter space was restricted by an ionization threshold and a lower bound on the non-linear phase shift, $\Delta\phi_{NL}^{min}$ [48], ensuring the simulated ensemble remained within the regime of valid experimental operation.

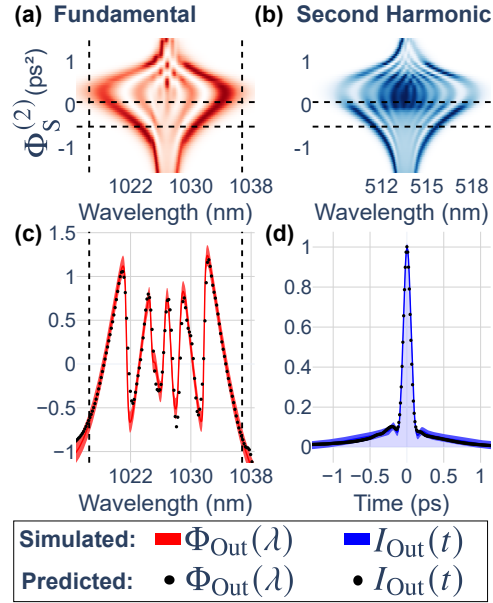

Fig. 2. **RD-scan retrieval of simulated pulse:** (a) Fundamental and (b) second-harmonic spectra of the simulated RD-scan. Black dashed lines in top row mark the limited region of $\Phi_S^{(2)}$ scan passed to the RD-scan for pulse prediction. (c) Phase $\Phi_{Out}(\lambda)$ of the simulated pulse (red) and RD-scan prediction (black) in radians. (d) Pulse-shape intensity $I_{Out}(t)$ of the simulated pulse (blue) and RD-scan prediction (black). Simulated targets are displayed with a vertical span of $\pm 120$ mrad and $\pm 2\%$ $I_{peak}$, respectively, for comparison with error ranges demonstrated in Fig. 3.

This NN architecture was built to process variable-length spectral scans and to support multiple output retrieval modes ($\Phi_{Out}$, $I_{Out}$, $\Phi_S$), enabling independent retrieval tasks, inter-nal self-consistency metrics, and pulse shaper optimization. A complete description of the simulations and the RD-scan transformer is provided in the Supplementary sections 2 and 3 respectively.

We must emphasize that, while we were able to closely reproduce the RD-scan spectrogram with Luna, perfect agreement is not necessary, instead a large physically realizable parameter space is the main requirement to train a predictive model [55, 56].

In Fig. 2 RD-scans predictions for the phase $\Phi_{Out}(\lambda)$ [panel (c)], $I_{Out}(t) \equiv |E(t)|^2$ [panel (d)] are presented for a simulated validation RD-Scan. Panels (a) and (b) show the simulated fundamental and second harmonic spectra as a function of RD-scan dispersion $\Phi_S^{(2)}$. The black dashed lines in (a) and (b) indicate the portion of the $\Phi_S^{(2)}$ scan passed to the NN for prediction; the NN has no knowledge of the $\Phi_S^{(2)}$ scan outside these limits.

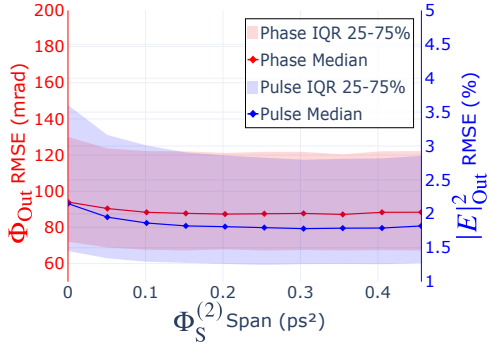

Fig. 3. **RD-scan performance across simulations:** Performance of RD-scan NN on 100k validation simulated pulses like that shown in Fig. 2. Error of RD-scan prediction vs simulated target is presented as a function of $\Phi_S^{(2)}$ scan length for phase $\Phi_{Out}(\lambda)$ and pulse shape $I_{Out}(t)$. The median value of phase error (spectral intensity weighted RMSE) is presented in red (left axis). The median value of pulse shape error (RMSE of $I(t) > 0.1\% I_{peak}$) is presented in blue (right axis). Inner quartile range, $25\% - 75\%$ of RMSE results in phase and pulse shape error are given for phase (pink) and pulse shape (light blue) respectively.

In Fig. 3, the performance of the RD-scan is measured for various scan lengths (1-10 spectra) spanning $\Phi_S^{(2)} = 0 - 0.45 ps^2$. For each scan length, the performance of the RD-scan network was measured across 10k simulated validation pulses. Remarkably, the median error of an RD-scan from a full $\Phi_S^{(2)} = 0.45 ps^2$ scan is only 15% lower $I_{Out}(t)$ RMSE than that of a single spectrum (1.82% vs 2.15% $I_{Peak}$ RMSE).

It is well established that mathematically, a single one-dimensional input does not hold enough information to uniquely characterize a pulse based on the Fourier transform [57–59]. Nevertheless, Fig. 3 empirically demonstrates high precision retrieval of realistic pulses from a single measurement (fundamental and second harmonic) over our extensive simulated parameter space. Although the retrieved pulse from a single measurement is not unique, across our realistic simulated bounds, the RD-scan achieves a high degree of precision in pulse retrieval from a single measurement. We attribute this capability to the network's learned extraction of phase information encoded in SPM spectral features and the high quality of the training set based on realistic simulations.

## 3. Experimental data comparison

Trained on simulated data, the RD-scan model was applied to experimental pulse retrieval. Fig. 4 shows a representative result using the same plotting elements as Fig. 2: the measured RD-scan fundamental (a) and second-harmonic (b) spectra (identical colormap), with RD-scan retrieved phase $\Phi_{Out}$ (c) and intensity $I_{Out}$ (d) overlaid in black on frequency-resolved optical gating (FROG) retrievals.

The example shown in Fig. 4 is one of twelve RD-scan pulse retrievals obtained from distinct pulses within a $\Phi_S^{(2)}$ scan spanning a factor-of-two range in $I_{Out}(t)$ FWHM duration (170–340 fs). When compared with independent SHG-FROG measurements, the RD-scan exhibits a mean FWHM disagreement of 22% for single-measurement retrievals and 17% for the full (0.45 ps$^2$) RD-scan. Both single-measurement and full-scan RD-scan retrievals accurately identify the minimum $\Phi_S^{(2)}$ location within the scan, with sub-2% FWHM disagreement relative to SHG-FROG. In addition, both RD-scan modalities exhibit substantially greater continuity in the retrieved FWHM across the scan than SHG-FROG.

We find that these properties, combined with the near-instantaneous pulse retrieval enabled by the transformer neural network, are well suited for routine pulse characterization and optimization. To this end, we have developed RD-scan software that continuously retrieves the pulse shape $I_{\text{Out}}(t)$ directly from spectrometer measurements. This live-view capability enables rapid, high-precision optimization by providing autocorrelator-like feedback in a substantially smaller form factor. In practice, RD-scan is alignment-insensitive and effective for fine-tuning pulse compression and for monitoring and quantifying pulse stability, including ionization-induced fiber instabilities.

## 4. Conclusion

We have introduced RD-scan, a transformer neural network pulse-retrieval technique that uses a pulse shaper to scan the spectral phase $\Phi_S^{(2)}$ prior to CPA and SPM while measuring the paired fundamental and second-harmonic spectra. We have validated the technique using a high average and peak power laser operating at 300 W, 1.2 ps, with a tunable rate from 25 to 200 kHz, and compressed >10x to 100 fs in a HCF. RD-scan performs robustly across a broad simulated parameter space, encompassing the operational conditions of our laser system, including variations in laser pulses, fiber geometries, and pulse compression. Evaluating 100 000 unique simulated validation pulses generated with `Luna` with varying $\Phi_S^{(2)}$ scan ranges, we find that even a single measured spectrum of the fundamental and the second harmonic encodes sufficient information for precise pulse retrieval. Finally, we experimentally demonstrate phase and pulse retrieval with RD-scan on both two-dimensional (2D) and one-dimensional (1D) measurements, enabling practical live-view pulse-characterization.

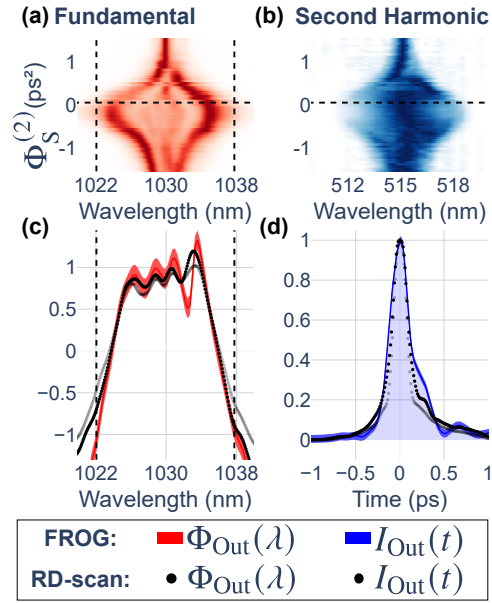

Fig. 4. **RD-scan retrieval of experimental pulse:** (a) Fundamental and (b) second-harmonic spectra of experimental RD-scan. Black horizontal line on top row indicates the $\Phi_S^{(2)}$ point of measurement. (c) Phase $\Phi_{\text{Out}}(\lambda)$ of independent FROG (red) and RD-scan (black) retrievals in radians. (d) Pulse-shape intensity $I_{\text{Out}}(t)$ from independent FROG (blue) and RD-scan (black) retrievals. On the bottom row black curves denote recovery using a 0.45 ps² RD-scan, while gray curves denote recovery from a single measurement. Vertical spans of FROG retrieval are identical to Figs. 2 to aid comparison.

## 5. Acknowledgments

KW was partially funded by AFOSR grant FA9550-21-1-0387. TS was partially funded by the US Department of Energy, Office of Science, Chemical Sciences, Geosciences, & Biosciences Division grant DE-SC0024508 to assist with pulse characterization, data analysis, and manuscript preparation. TCW where funded by DE-FG02-08ER15984. The laser system was funded by ONR grant N00014-19-1-2339 and by the College of Liberal Arts and Sciences at the University of Connecticut. We thank K. Bodek for insightful discussions.

**Data availability:** The RD-scan simulation and neural network training code are available at https://github.com/uconnultrafast/Remote-Dispersion-Scan-Simulation-and-Neural-Network-Training (v1.0.0). The data presented in this paper may be obtained from the corresponding author upon reasonable request.

**Disclosure:** The authors declare no conflicts of interest.

# Supplement: Remote dispersion scan: transformer-network retrieval of ultrafast pulses after nonlinear propagation

KEVIN WATSON[1], YUTONG GENG[1], TOBIAS SAULE[1], KOVI BODEK[2], THOMAS WEINACHT[2], AND CARLOS A. TRALLERO-HERRERO[1,*]

[1]*Department of Physics, University of Connecticut, Storrs, CT 06269, USA*
[2]*Department of Physics and Astronomy, Stony Brook University, Stony Brook, New York 11794, USA*
[*]*Corresponding author: carlos.trallero@uconn.edu*

## 1. RD-scan Internal Layout

A remote dispersion scan (RD-scan) operates by employing an internal pulse shaper to systematically control the spectral phase of the laser pulses prior to nonlinear propagation. It is important to note that, in our system—similar to many ultrafast laser platforms with integrated pulse shaping—the $4f$ pulse shaper is positioned before the main amplification stage rather than after it. Consequently, the applied phase modulation precedes both amplification and compression within the laser system.

Figure S1 illustrates the placement of the internal pulse shaper relative to the major components of the laser architecture. To validate the accuracy of the numerical model, simulated RD-scan results were directly compared with experimental scans acquired using the internal pulse shaper. This comparison confirms that the simulations reliably capture the cumulative effects of phase modulation applied by the pulse shaper, including subsequent amplification, spectral broadening, and nonlinear propagation through the hollow-core fiber.

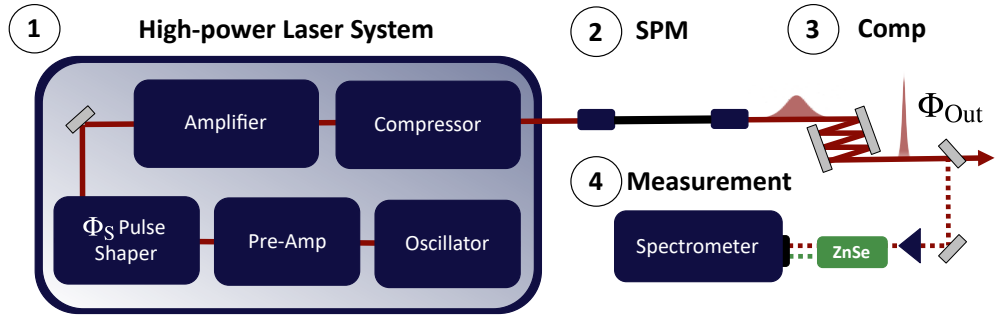

Fig. 1. **Internal layout of the remote dispersion scan (RD-scan) system.** The RD-scan methodology consists of four primary stages: (1) a laser system incorporating an internal pulse shaper; (2) spectral broadening via nonlinear propagation in a hollow-core fiber; (3) pulse compression using a sequence of chirped mirrors; and (4) spectral characterization of the pulse through second-harmonic generation (SHG). In this figure, Stage (1) is expanded to illustrate the principal components of the high-power laser system, including pulse generation in the oscillator, pre-amplification to the desired power level, spectral phase modulation using a $4f$ pulse shaper, the main amplification stage, and subsequent pulse compression.

## 2. RD-scan Simulations

At the core of the RD-scan approach is a neural network (NN) that accurately recovers a pulse by predicting the nonlinear effects of self-phase modulation (SPM) encoded in the broadened spectra. NNs are fundamentally different in design from classical algorithmic approaches [1–4]: NN architectures are trained from data rather than hand-engineered to perform a specific task [5–7].

Data used to train neural networks may be generated in two ways: simulated or experimental [8]. In our case the datasets comprise pulse characterizations (phase and pulse shape) together with the corresponding fundamental and second-harmonic spectra. Experimental data are advantageous because they incorporate all real nonlinear effects and instrument-specific nuances (for example, spectral absorption and spectrometer resolution) directly into the measurements used for training. However, this advantage can also be a liability: if any instrument is replaced, the resulting change in the measurement chain can permanently degrade a model's performance. The experimental route is also time intensive when extensive coverage of the pulse parameter space is required. Simulated data is quick to generate and therefore easier to modify facilitating the correction or emulation of experimental effects. Because the simulation code developed for this work is made available, future users of the RD-scan method can readily adapt the simulated parameter space to accurately match their specific experimental conditions. In practice, RD-scan can be implemented for an individual laser system by changing a few simulation variables to mach there system (for example, transform-limited (TL) pulse duration, center wavelength, and spectrometer resolution) to create a customized training dataset.

The current parameter space of data used to train the neural network presented in this paper spans a nine-dimensional parameter space. This space was selected to account for variations in the laser pulse, hollow-core fiber, and pulse compression. Our simulations start with a given input pulse coupled into a hollow-core fiber. The parameter of pulse energy accounts for not only degradation or power fluctuation of our laser but also fluctuations in coupling efficiency to our hollow-core fiber. The phase of our pulse also plays a key role, as phase determines the pulse shape and peak intensity that drive SPM and self-steepening. To account for a range of phases, we use a Gaussian pulse and alter its phase using a Taylor expansion of phase terms to the fourth order ($\Phi_S^{(2)}$, $\Phi_S^{(3)}$, $\Phi_S^{(4)}$). A parameter space of the fiber is also added, with the most critical being fiber pressure. By training the RD-scan network on a wide range of pulse energies and fiber pressures, we train RD-scan over a range of operational conditions we use for a wide range of repetition rates (in our case from $25\,\text{kHz}$–$100\,\text{kHz}$). Additionally, fiber length and diameter are altered such that the same RD-scan neural network can perform on a wide range of different fibers without the need for retraining. Finally, the level of compression after SPM is included in our parameter space; training over a range of compression is critical, as it allows the model to learn how variously compressed pulses respond. We can then make changes to our beamline (adding chirped mirrors or windows, e.g.) with confidence, knowing the network is capable of retrieval over a wide range of pulse compressions.

In Table 1, the specific ranges of simulation are given.

Fiber propagation was simulated with the state-of-the-art nonlinear capillary and fiber propagation package `Luna.jl`, which accounts for a wide range of effects including multimode propagation and the Kerr nonlinearity. For computational efficiency, we employ bimodal propagation [9]: the intensity in the third mode was observed to contribute less than $5.00\,\%$ of the total power, while the computational cost increases linearly with the number of modes.

After propagation, we apply second-order dispersion compensation and generate a set of 100 second-harmonic (SHG) spectra corresponding to different degrees of pulse compression. The same simulation is repeated in parallel to form a full dispersion scan by varying the pre-SPM dispersion $\Phi_S^{(2)}$ over the range $-1.50\,\text{ps}^2$ to $1.50\,\text{ps}^2$. Each complete simulated dispersion scan is stored as an `HDF5` dataset; each dataset samples a unique parameter space point, whose ranges are summarized in Table 1.

Pulse energy and gas pressure were sampled in two nested loops over ten evenly spaced values each, so that the dataset uniformly covers the operational regimes of the various repetition rates considered. To restrict the database to physically realistic pulses that have undergone significant SPM, we implement two prescreening boundary conditions. Parameter combinations that do not satisfy these criteria are skipped. The prescreening criteria are:

Table 1. Simulation Parameter Space

| Parameter | Range / Values |
|---|---|
| Pulse energy | $\mathcal{E}$ = 1 mJ – 6 mJ |
| Phase applied by pulse shaper | $\Phi_S^{(2)}$ = –0.75 to 0.75 ps$^2$ |
| | $\Phi_S^{(3)}$ = –0.2 to 0.2 ps$^3$ |
| | $\Phi_S^{(4)}$ = –0.065 to 0.065 ps$^4$ |
| Fiber pressure (Kr) | $p$ = 500 mBar – 2.5 Bar |
| Fiber diameter | $d$ = 0.6 mm – 0.8 mm |
| Fiber length | $l$ = 1 m – 5 m |
| Compression applied after SPM | $\Phi_C^{(2)}$ = –5k to –15k fs$^2$ |
| | $\Phi_C^{(3)}$ = –1k to 1k fs$^3$ |

- the peak intensity ($I_{Peak}$) remains below the ionization threshold of the fill gas (Kr)
- the accumulated nonlinear phase satisfies $\Phi_{NL} \geq 2.5\pi$.

Associated metadata (pulse energy, gas pressure, fiber length, core diameter, and dispersion-scan parameters) are stored with each `HDF5` entry to facilitate clear documentation.

## 3. Neural Network Training

Once a representative dataset has been created, the data are passed through a pretraining data augmentation stage. The goal of this step is to accurately modulate the simulated data to match experimental measurements. These steps include applying Gaussian smoothing to mimic the resolution of the experimentally used spectrometer. The center wavelengths of both the fundamental and second-harmonic spectra are then shifted such that the neural network learns to extract information from the spectral shapes of the fundamental and second harmonic rather than from their relative positions. This procedure additionally improves robustness to small variations in center wavelength and to errors in spectrometer calibration.

Finally, a point in the RD-scan is selected as the target, and a sequence of $N$ spectra is extracted from the full simulated scan for grouping into training batches. The choice of a target point in the scan that is not centered at zero reflects two important considerations. First, it allows the model to learn from pulses that possess a large amount of second-order dispersion prior to SPM. Second, it increases the number of available target pulses for training by the number of distinct scan positions that can be selected; in our case, this results in approximately a 40× increase in the effective training set size. Each time a group of length-$N$ scans is extracted from the simulated dataset and placed into a batch, a new random target is selected from within the defined target window of the scan.

Once the data have undergone preprocessing and batching, the neural network is trained across all batches for a single epoch, with the full training procedure repeated between 50 and 200 epochs. Throughout this learning process, Gaussian noise is dynamically added to both the fundamental and second-harmonic spectra within each batch to further bridge the gap between simulated data and experimental measurements. The noise level is augmented from a standard deviation of 2% down to 0% of the peak intensity of each individual spectrum, with the fundamental and second-harmonic spectra evaluated separately. Training is initialized at the highest noise levels such that broad spectral features are captured early in the optimization process. The noise amplitude is then gradually reduced, ultimately reaching zero during the final epoch, allowing the model to resolve finer details without relying excessively on them for retrieval. In addition, weight decay is applied such that all network weights are progressively driven toward smaller values in proportion to their magnitude. A dropout rate of 0.05% is also employed across all layers, randomly setting 0.05% of the weights to zero during training. Together, weight decay and dropout act as key regularization mechanisms that promote robust model weighting, encouraging reliance on a large number of redundant perceptrons and thereby improving the stability and consistency of pulse retrieval [10].

Once the model makes a predictions for a given batch there must be a numeric evaluation of the quality of performance of the neural network to the known properties of the pulses from our data. This numeric evaluation of error is given by a loss function. For our loss function we implemented a mean squared error loss as given by the equation bellow:

$$\text{Loss} = \text{Mean}\left(\frac{\sum\left[(\text{pred} - \text{target}) \cdot \text{mask}\right]^2}{\sum \text{mask}}\right) \tag{1}$$

The loss function was selectively masked according to the prediction target associated with each learned query. The phase loss was weighted by the peak-normalized spectral intensity of the target pulse. The pulse-shape loss employed a binary temporal mask defined by the simulated target pulse intensity threshold $I(t) > 0.01\% I_{\text{peak}}$. In contrast, prediction of the quadratic phase parameter $\Phi_s$ required no masking, as it is a scalar. A single model was trained to predict all three queries simultaneously by summing their respective loss terms and backpropagating the combined loss. The individual losses were not equally weighted: the phase, pulse-shape, and $\Phi_s^2$ losses were scaled by coefficients [1, 1, 0.001], ensuring balanced optimization of phase and pulse shape while significantly down-weighting the contribution of the quadratic phase term.

By training the model to predict both the spectral phase and the time-domain pulse shape, we obtain two complementary representations of the same underlying electromagnetic field. Because the two representations $\Phi(\omega)$ and $I(t)$ of a pulse are related by the Fourier transform for a known spectra $S(\omega)$.

$$E(t) = \frac{1}{2\pi} \int_{-\infty}^{\infty} \sqrt{S(\omega)}\, e^{i\Phi(\omega)}\, e^{-i\omega t}\, d\omega, \tag{2}$$

$$S(\omega) = \left|\tilde{E}(\omega)\right|^2 \qquad I(t) = \left|E(t)\right|^2 \tag{3}$$

This redundancy of $\Phi(\omega)$ and $I(t)$ learned queries provides a self-consistency metric that strengthen retrieval confidence and alerts us to unphysical retrievals. We quantify the agreement between the two reconstructions using a normalized $\ell_2$ disagreement metric. Denoting the time-domain pulse shape prediction as $I_{\text{PS}}(t)$ and the pulse reconstructed from the predicted spectral phase as $I_\Phi(t)$, we define

$$\text{CI} = \frac{\left(\sum_{k=1}^{N} [I_{\text{PS}}(t_k) - I_\Phi(t_k)]^2\right)^{1/2}}{\left(\sum_{k=1}^{N} [I_{\text{PS}}(t_k)]^2\right)^{1/2}}, \tag{4}$$

where CI is an internal confidence metric: small values indicate strong self-consistency (high confidence in the retrieval), while large values indicate disagreement and suggest the predicted pulse may not be physically consistent with the measured spectrum. Because the network is trained on physically consistent pulses, this metric serves both as an uncertainty proxy and as a diagnostic for nonphysical or out-of-distribution predictions.

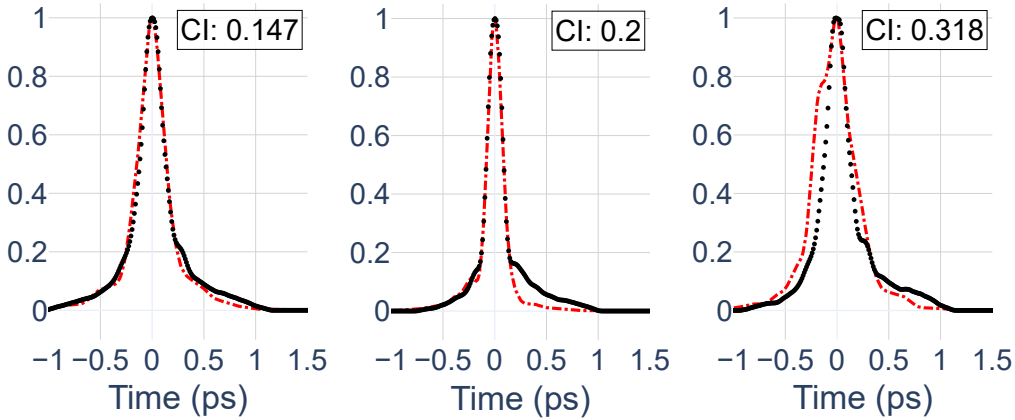

Fig. 2. **Internal confidence demonstration.** Representative time-domain pulse retrievals obtained with the RD-scan network for three experimental pulses, each annotated with the network's internal confidence index (CI) in the upper-right corner of the panel. The direct time-domain retrieval $I(t)$ is shown in black; the red trace is the reconstruction produced by applying the network-predicted spectral phase to the measured spectrum followed by an inverse Fourier transform. The panels illustrate typical behaviors observed in the dataset: the example with CI = 1.5 (and other measurements at comparable CI values) exhibits near-perfect overlap between direct and reconstructed traces, indicating very high retrieval fidelity; the panel with CI = 0.2 shows strong overall agreement with moderate residual discrepancies; the panel with CI = 0.3 displays pronounced waveform distortion, consistent with a failed or unreliable retrieval. These examples demonstrate that the CI metric correlates with reconstruction physical consistency and can be used to identify high-confidence versus low-confidence RD-scan retrievals.

## 4. Model architecture and motivation

Recent advances in neural networks have shown that common architectures (e.g., fully connected, convolutional, transformer) can learn a broad range of image processing tasks when provided with a large amount of data and compute [11–15]. However, architectural choices in how data are represented and processed strongly affect training efficiency, inference speed, and the model's versatility [12–16]. For the RD-Scan pulse-retrieval problem we adopt a transformer-based architecture because of its natural ability to process variable-length sequences of tokens (here: spectral columns) and to support multiple learnable query vectors [3] that drive independent retrieval tasks and internal self-consistency checks as shown in the pervious section.

A transformer based neural-network (TNN) architecture was built specifically for RD-scan pulse retrieval, and details of its configuration are given in S2. In this design, each spectrum is partitioned into its fundamental and SHG components which are interleaved as tokens, enabling the TNN to attend to the cross-modal relationships between the fundamental and SHG signals throughout the scan. Transformer encoders are desirable for their ability to process any length of scan and transformer decoder's can support multiple learnable queries for each scan. Learnable queries are input vectors passed to the decoder that act as requests specifying which physical quantity the network will reconstruct. In this work, the TNN was trained to output three learnable queries: the spectral phase of the pulse, $\Phi_{\text{Out}}(\lambda)$, the temporal pulse shape, $I_{\text{Out}}(t)$, and the Group delay dispersion of the pulse before entering the fiber, $\Phi_S$. This enables fast, on-demand retrieval of all three quantities. The discussion and presentation of $\Phi_S$ retrieval through learned query has been moved to the supplementary but is of particular note as accurate $\Phi_S$ retrieval demonstrates that the RD-scan method is capable of extrapolating information from points prior to the location of spectral measurement.

The complete model used for RD-Scan is illustrated in Fig. S3. Input data are shown at the base of the diagram and the forward data flow proceeds upward to the outputs. Each processing block is labeled with its component name and operational details. A separate panel reports model statistics (total parameter count and the parameter distribution across the encoder, transformer layers, and encoder–decoder MLPs). Rather than using a pure transformer encoder–decoder stack, we interpose feed-forward multilayer perceptron (MLP) blocks before and after the transformer encoder and decoder; these MLPs transform local spectral features and provide the necessary projection/integration stages between token encoding and query-driven decoding (see Fig. S3 for the exact block diagram and layer sizes). Taken together, this design combines the transformer's flexible sequence modeling with task-specific MLP projections and multiple learned queries, producing a compact, efficient architecture tailored for high-fidelity, self-consistent pulse retrieval. We note that the overall number of model parameters was deliberately matched to the size of the simulated dataset. This ensures sufficient capacity to accurately capture the physical effects represented in the simulations, while avoiding excessive flexibility that could lead to overfitting and degraded performance on experimental data.

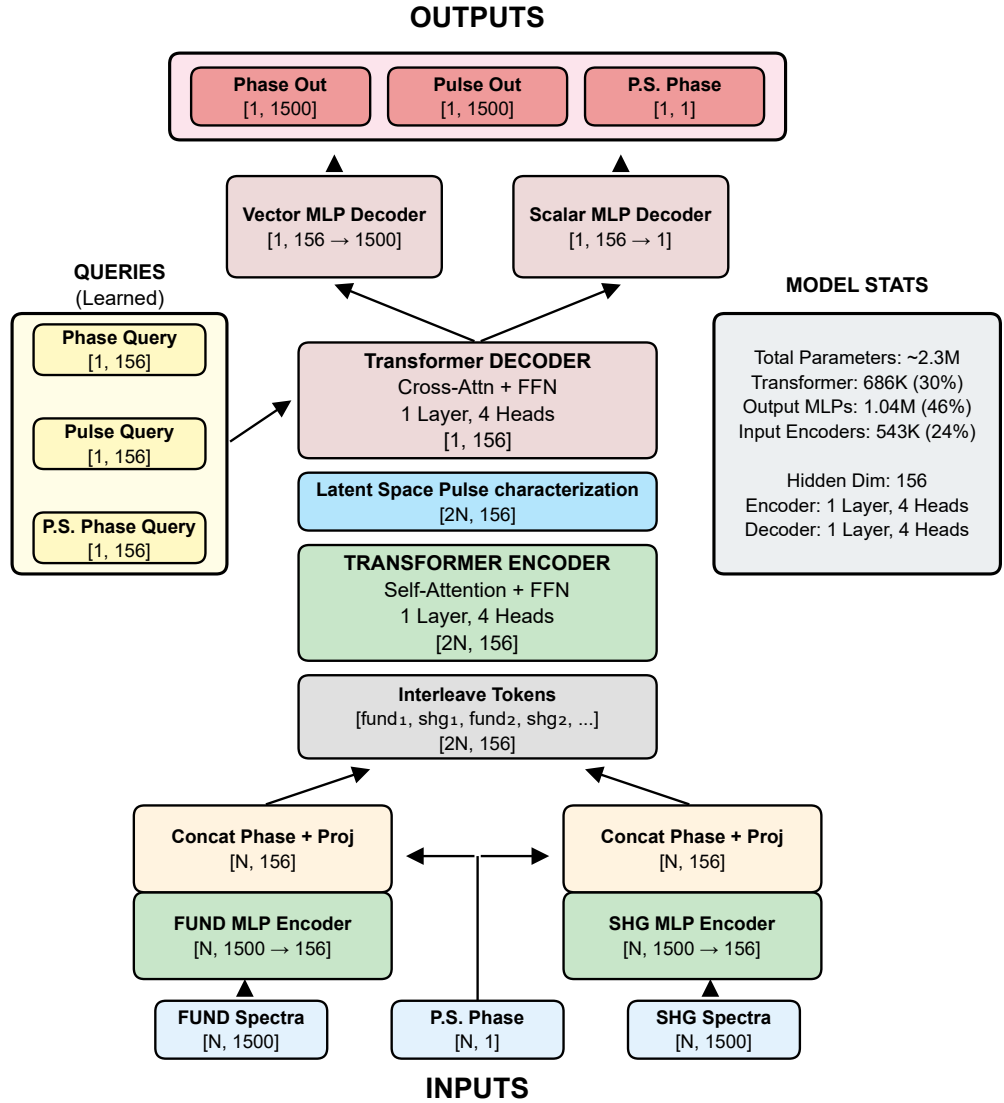

Fig. 3. **Full architecture of the transformer-based neural network used for RD-Scan pulse retrieval.** The network combines feed-forward and transformer layers to process spectral data. Fundamental and second-harmonic (SHG) encoders downscale the input spectra into latent-space representations, which are then appended with the relative pulse shaper phase $\Phi_S^{(2)}$. These representations are passed to the transformer encoder, producing a latent-space characterization of the pulse. The transformer decoder combines this encoded information with a user-specified learned query, and the decoder multilayer perceptrons scale the resulting output to the target vector length. This design enables simultaneous retrieval of multiple complementary pulse properties while supporting internal self-consistency checks.

## 5. Simulation Results

We evaluated RD-Scan on simulated data not used in training (validation) generated within the same parameter space outlined in 1; the results are shown in Fig. 3 and demonstrate robust retrieval performance using a small number of spectra. Here $N$ denotes the number of spectra sampled at regularly spaced intervals of $\Phi_S^{(2)}$; each reported value of $N$ is computed over 10,000 independent validation examples. For each learned query we report both the sample mean and the median so that readers may judge the influence of outliers on central tendency. Overall performance is further summarized by the interquartile range (IQR), which displays the 25th and 75th percentiles and thereby characterizes the typical spread of retrieval errors across the validation set.

For completeness, Table 3 reports the network performance for the learned $\Phi_S$ query (omitted from Fig. 3 for clarity). This query retrieves the group-delay dispersion prior to SPM and CPA. The network reliably recovers $\Phi_S^{(2)}$ with a relative error as low as 0.319 for simulated pulses randomly sampled from the range $[-0.5\,\text{ps}^2, 0.5\,\text{ps}^2]$. This result demonstrates the model's ability to predict pulse information that is remote from the spectral measurement point. Accurate phase retrieval prior to self-phase modulation (pre-SPM) is particularly valuable for closed-loop optimization of output pulses via feedback to the lasers integrated pulse shaper.

Table 2. Performance statistics of phase $\Phi_{Out}$ and $I_{Out}$ pulse output retrieval as shown in Fig. 3 presented in tabular format. Span of $\Phi_S^{(2)}$ dispersion scan is presented as a function of the number of spectra $N$, where each spectra is separated by $0.05 ps^2$. Each row reports the mean, median, and interquartile range (IQR 25% – 75% percentage interval). All values for phase are given in units of radians of intensity weighted RMSE of each validation sample. All units of Pulse are given in % peak intensity of RMSE of $I(t) > 0.01\% I_{peak}$.

| | Phase | | | Pulse | | |
|---|---|---|---|---|---|---|
| N | Mean | Median | IQR (25%-75%) | Mean | Median | IQR (25%-75%) |
| 1 | 144 | 94.0 | 58.0 (72.1–130) | 3.40 | 2.15 | 2.15 (1.45–3.60) |
| 2 | 135 | 90.4 | 54.8 (68.9–124) | 3.18 | 1.95 | 1.82 (1.34–3.16) |
| 3 | 134 | 88.4 | 54.6 (67.6–122) | 3.11 | 1.86 | 1.72 (1.29–3.01) |
| 4 | 131 | 87.8 | 54.9 (67.2–122) | 3.06 | 1.82 | 1.63 (1.28–2.91) |
| 5 | 131 | 87.4 | 53.3 (67.8–121) | 3.06 | 1.81 | 1.60 (1.26–2.86) |
| 6 | 130 | 87.6 | 54.6 (67.2–122) | 2.98 | 1.80 | 1.58 (1.25–2.83) |
| 7 | 132 | 87.7 | 54.3 (67.5–122) | 2.97 | 1.78 | 1.54 (1.26–2.80) |
| 8 | 126 | 87.2 | 53.2 (67.2–120) | 2.94 | 1.79 | 1.55 (1.26–2.81) |
| 9 | 133 | 88.3 | 54.7 (67.4–122) | 2.97 | 1.79 | 1.56 (1.26–2.82) |
| 10 | 131 | 88.4 | 55.0 (67.2–122) | 3.07 | 1.82 | 1.58 (1.27–2.86) |

Note: $N_{\text{samples}} = 10^4$ for every row.

Table 3. $\Phi_S$ RD-scan learned query retrieval performance of group delay dispersion prior entering nonlinear propagation $\Phi_S$ as a function of the number of spectra $N$ on simulated validation pulses. Displayed values are rounded to 3 significant figures.

| | $\Phi_S$ | | |
| --- | --- | --- | --- |
| N | Mean | Median | IQR (25%-75%) |
| 1 | 1.26 | 0.654 | 0.788 (0.310–1.10) |
| 2 | 1.13 | 0.431 | 0.665 (0.190–0.860) |
| 3 | 1.02 | 0.358 | 0.621 (0.160–0.780) |
| 4 | 0.973 | 0.345 | 0.584 (0.150–0.730) |
| 5 | 0.990 | 0.344 | 0.603 (0.150–0.750) |
| 6 | 1.00 | 0.333 | 0.587 (0.150–0.730) |
| 7 | 0.961 | 0.322 | 0.564 (0.140–0.710) |
| 8 | 0.968 | 0.323 | 0.585 (0.140–0.730) |
| 9 | 0.972 | 0.324 | 0.581 (0.150–0.730) |
| 10 | 0.947 | 0.319 | 0.570 (0.140–0.710) |

Note: $N_{\text{samples}} = 10^4$ for every row.